\documentclass[prd,preprint,superscriptaddress,nofootinbib,longbibliography,aps]{revtex4-1}

\usepackage[utf8]{inputenc}
\usepackage{graphicx}
\usepackage{mathtools}
\usepackage{longtable}
\usepackage{amsmath}
\usepackage{tikz-feynman}
\usepackage{color} 
\usepackage{amssymb}
\usepackage{subfigure}
\usepackage{microtype}
\usepackage[linktoc=all]{hyperref}
\usepackage{cleveref}
\usepackage{stackengine}
\usepackage{color}
\usepackage{bm}
\usepackage{braket}
\usepackage{ulem,xpatch}
\usepackage{wasysym}
\usepackage{slashed}
\usepackage{array}
\usepackage{graphicx}
\usepackage{tikz}
\usetikzlibrary{calc,arrows.meta}
\usepackage{hyperref}

\usepackage{lipsum}
\usepackage{setspace}
\usepackage{etoolbox}
\makeatletter
\patchcmd{\@footnotetext}
  {\setspace@singlespace}{1}
  {}{}
\makeatother

\def\beq{\begin{eqnarray}}
\def\eeq{\end{eqnarray}}

\newcommand{\vev}[1]{ \left\langle {#1} \right\rangle }

\def\SU{\mathop{\rm SU}}
\def\U{\mathop{\rm U}}

\def\ten{\mathbf{10}}
\def\fives{\bar{\mathbf{5}} }
\def\five{\mathbf{5}}
\def\Torb{\mathbf{T}^2/\mathbb{Z}_3}
\def\Ztwo{\mathbb{Z}_2}
\def\Htwo{H_{\mathrm{I\!I}}}
\def\Hthr{H_{\mathrm{I\!I\!I}}}

\begin{document}  
\title{Three-zero texture of quark-mass matrices as a solution to the strong CP problem }

\author{Qiuyue Liang}
\email{qiuyue.liang@ipmu.jp}
\affiliation{Kavli Institute for the Physics and Mathematics of the Universe (WPI), University of Tokyo, Kashiwa 277-8583, Japan}

\author{Risshin Okabe}
\email{risshin.okabe@ipmu.jp }
\affiliation{Kavli Institute for the Physics and Mathematics of the Universe (WPI), University of Tokyo, Kashiwa 277-8583, Japan} 

\author{Tsutomu T. Yanagida}
\email{tsutomu.tyanagida@sjtu.edu.cn }
\affiliation{Kavli Institute for the Physics and Mathematics of the Universe (WPI), University of Tokyo, Kashiwa 277-8583, Japan} 
\affiliation{Tsung-Dao Lee Institute \& School of Physics and Astronomy, Shanghai Jiao Tong University, China}

\begin{abstract}
The strong charge-parity (CP) problem has been a long-standing problem in particle physics since 1976, illustrating the small CP-violation phase in quantum chromodynamics (QCD). The axion, based on the Peccei-Quinn mechanism, is the most popular solution to the problem. In this paper, we propose an alternative solution based on the three-zero texture of quark mass matrices without additional heavy quark states, which has been shown to fit data well. We show that the required three-zero texture is naturally constructed in a six-dimensional spacetime with a $\mathbf{T}^2/\mathbb{Z}_3$ orbifold compactification.
\end{abstract}

\maketitle

\section{Introduction}
\label{sec:intro}

The strong charge-parity (CP) problem has been a long-standing issue in particle physics since 1976 \cite{tHooft:1976snw}. Without additional symmetry, the CP-violating angle in quantum chromodynamics (QCD), $\bar{\theta}$, should have an $\mathcal{O}(1)$ value. However, the current upper limit from the neutron electric dipole moment \cite{Abel:2020pzs} implies an extremely suppressed CP-violating parameter, $|\bar{\theta}| \lesssim 10^{-10}$. The axion \cite{Weinberg:1977ma,Wilczek:1977pj}, based on the Peccei-Quinn mechanism \cite{Peccei:1977hh}, is the most popular solution to this problem, potentially originating from high-energy fundamental physics such as string theories \cite{Svrcek:2006yi,Burgess:2023ifd}. However, experiments have not identified such particles yet. Therefore, it remains important to construct alternative solutions to the strong CP problem. 

Some alternative solutions have been proposed based on spontaneous CP violation. The CP is assumed to be an exact symmetry at the fundamental level which is spontaneously broken at some energy scale. In these types of models, the QCD vacuum angle vanishes,  $\theta_0 =0$, before the spontaneous CP violation. However, spontaneous CP breaking induces complex phases in up-type and down-type quark mass matrices $M_u$ and $M_d$ to explain the observed CP-violating phase in the Cabibbo--Kobayashi--Maskawa (CKM) quark mixing matrix, which generates a shift of the vacuum angle. The physical vacuum angle $\bar{\theta}$ is obtained through
\begin{equation}
    {\bar \theta} = \theta_0 + \text{Arg} [\text{det}(M_d) \text{det} (M_u)] \ ,
\end{equation} 
which is the predicted CP-violating angle that is proportional to the neutron electric dipole moment. Therefore, the question is how to control the quark mass matrices by imposing additional symmetries so that ${\bar \theta}$ remains vanishing.

It is clear that we have ${\bar \theta} = 0$ if the quark mass matrices are Hermitian. This can be easily realized by introducing the horizontal gauge symmetry $\SU(3)_H$ \cite{Masiero:1998yi,Choi:2019omm}. However, to extend this to the lepton sector, it is necessary to introduce second-rank tensor (sextuplet) Higgs bosons, which might violate the Hermitian nature of the quark matrices at higher orders.

An alternative approach to obtaining a real determinant is to ensure that the complex matrix elements always encounter zero-valued elements when the determinants of the mass matrices are calculated. This is the Nelson-Barr mechanism \cite{Nelson:1983zb,Barr:1984qx}, which assumes an extra pair of heavy quarks. The key point is that all CP-violating phases are located in the off-diagonal elements between the heavy quarks and the Standard Model quarks, and these complex elements always interact with zero matrix elements when taking the determinant. Consequently, the determinant of the total quark mass matrices becomes real, and the physical CP-violating angle vanishes, ${\bar \theta}=0$. 

In this paper, we do not introduce additional heavy quark pairs but solve the strong CP problem by considering multi-zero textures of the quark mass matrices. Many-zero textures have a higher chance for the complex elements to hit the zeros in the determinant.\footnote{A similar idea has been recently proposed based on the modular invariance \cite{Feruglio:2024ytl,Petcov:2024vph}, and multi-Higgs~\cite{Hall:2024xbd}.} However, it is known that we can maximally have three zeros in the $3\times 3$ complex matrix to have at least one physical complex phase \cite{Harigaya:2012bw}. In \cite{Tanimoto:2016rqy}, it has been shown that there are 13 three-zero textures of $M_d$ among 20 possibilities which are consistent with observations by assuming $M_u$ to be diagonal. Among these thirteen textures, six of them have a real det$(M_d)$ despite the matrix being complex. Thus, the strong CP problem can be solved if the down-type quark mass matrix has the desired textures. However, such a three-zero texture still suffers from the fine-tuning problem, and we need to develop a mechanism to explain the zeros. In this paper, we attempt to construct a three-zero texture for $M_d$, with a diagonal up-type quark mass matrix, by imposing additional symmetries, particularly by extending to extra spacetime dimensions.

In Sec.\ref{sec:three-zero}, we discuss how imposing symmetry in 4D spacetime cannot resolve the problem, making it necessary to seek solutions in higher dimensions. In Sec.\ref{sec:extra-dim}, we discuss the orbifold compactification of higher dimensions and the additional $\Ztwo$ symmetry. We further construct the desired mass matrix and discuss the spontaneous breaking of CP symmetry. 

We use the $\SU (5)_\mathrm{GUT}$ convention, $\ten_i= (q_L, \bar u_R)_i$, and $  \fives_i = (\bar d_R)_i$, where $i=1,2,3$ denote family indices. We take all fermions to be left-handed for notation simplicity. We do not unify all standard-model (SM) gauge groups in the $\SU(5)_\mathrm{GUT}$, and do not discuss supersymmetry or superpartners in this paper.

\section{Construct three-zero textures in 4D}
\label{sec:three-zero}

In this section, we will demonstrate the difficulty of realizing such three-zero textures in the quark mass matrices by imposing symmetries in 4D. We use the first texture $M_d^{(1)}$\footnote{We will drop the superscript for simplicity in the remaining context.} in \cite{Tanimoto:2016rqy},
\begin{equation}
    M_d = \left(
        \begin{array}{ccc}
            0 & a & 0 \\
            a' & b e^{-i\phi} & c \\
            0 & c' & d
        \end{array}
    \right)\ ,
    \label{eq:Md1}
\end{equation}
as an example to illustrate this point. This matrix has been shown to be consistent with all the CKM angles and quark mass observations, which completely fix all seven real parameters, $(a, a', b, c, c', d, \phi)$, in the mass matrix, given in Table 2 of \cite{Tanimoto:2016rqy}.
We summarize those values in Tab.\ref{tab:params}.
Given this three-zero structure, one can see only real parameters $aa'd$ enters the determinant of the mass matrix, and physical vacuum angle, $\bar \theta  = 0$. 
 
\vspace{\baselineskip}
\begin{table}[h]
    \centering
    \begin{tabular}{|c|c|c|c|c|c|c|}
    \hline
       $a$ [MeV] & $a'$ [MeV] & $b$ [MeV] & $c$ [MeV] & $c'$ [GeV] & $d$ [GeV] & $\phi$ [$\circ$]  \\ \hline
       16\,-\,17.5 & 10\,-\,15 & 92\,-\,104 & 78\,-\,95 & 1.65\,-\,2.0 & 2.0\,-\,2.3 & 37\,-\,48 \\
       \hline
    \end{tabular}
    \caption{The allowed values of parameters of down-type quark mass matrix in Eq.\eqref{eq:Md1}, given in \cite{Tanimoto:2016rqy}.}
    \label{tab:params}
\end{table}
 
To construct this matrix with an additional symmetry in 4D spacetime, we first assume a global $\U(1)$ symmetry acting on quark families. We assign the $\U(1)$ charges for the quarks as $\xi_{i}$ for $\ten_{i}$ and $\xi'_{i}$ for $\fives_{i}$.
We set the charge of the Higgs $H$ to zero, as it can always be neutralized via the $\U(1)_Y$ gauge rotation. 
To have real $(M_{d})_{12}$ and $(M_{d})_{21}$, we need $\xi_1+\xi'_2=0$, and $\xi'_1+\xi_2=0$. This leads to $\xi_1+\xi'_1=-(\xi_2+\xi'_2)$. 

To recover the CP-violating phase in the SM, we introduce CP-violating singlet scalar fields $\eta_1$ and $\eta_2$. They share the same $\U(1)$ charges, $\xi_{\eta_1} = \xi_{\eta_2} = -(\xi_2 + \xi'_2)$, allowing for effective couplings $\ten_2 H \fives_2 \eta_{1,2}$, which make $(M_{d})_{22}$ complex by non-zero vacuum expectation values (VEVs) $\vev{\eta_{1,2}}$.\footnote{Notice that we need two $\eta$ fields because one of the phases can always be absorbed by a $\U(1)$ rotation.} 
Since $\xi_1+\xi_1' = -(\xi_2+\xi_2')$, we naturally obtain effective couplings $\ten_1 H \fives_1 \eta^\dagger_{1,2}$, resulting in a non-vanishing $(M_{d})_{11}$.  
No additional symmetry suppresses the coefficient for this term, so the determinant of $M_d$ is no longer real. This conclusion remains valid even when considering other symmetry groups or supersymmetry in 4D. To decouple $\eta^\dagger$ and $\eta'^\dagger$ from the first family quarks, we turn to extra spacetime dimensions, which we discuss in the next section.

\section{Engineering the three-zero texture in extra dimensions} 
\label{sec:extra-dim} 

In this section, we present a successful model in six-dimensional spacetime. We compactify two space dimensions to an orbifold, $\Torb$ \cite{Watari:2002fd}, where $\mathbf{T}^2$ stands for the 2-torus, and $\mathbb{Z}_3$ denotes three fixed points. Orbifold compactification is necessary to have chiral fermions in the four-dimensional spacetime. 
 
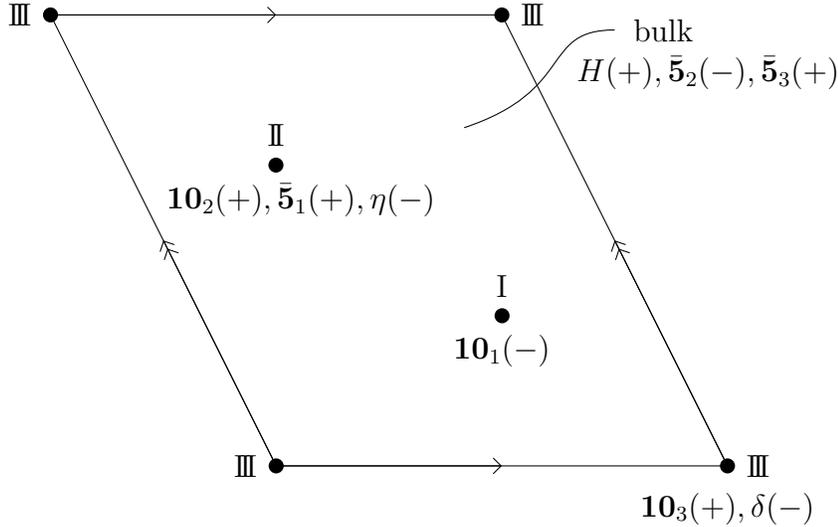
\begin{figure}
    \centering
    \begin{tikzpicture}[>={Straight Barb[length=3pt]}, inner sep=7]
        \coordinate[label=left:{I\!I\!I}] (O) at (0,0);
        \coordinate[label=right:{I\!I\!I}] (A) at (6,0);
        \coordinate[label=right:{I\!I\!I}] (B) at (3,6);
        \coordinate[label=left:{I\!I\!I}] (C) at (-3,6);
        \coordinate[label=above:{I}] (I) at ($(A)!0.333!(C)$);
        \coordinate[label=above:{I\!I}] (II) at ($(A)!0.667!(C)$);

        \draw (O) -- (A) -- (B) -- (C) -- cycle;
        \draw[->] (O) -- ($(O)!0.5!(A)$);
        \draw[->>] (A) -- ($(A)!0.5!(B)$);
        \draw[->>] (O) -- ($(O)!0.5!(C)$);
        \draw[->] (C) -- ($(C)!0.5!(B)$);

        \fill (O) circle [radius=0.1];
        \fill (A) circle [radius=0.1] node [below=0.1] {$\ten_3 (+), \delta (-)$};
        \fill (B) circle [radius=0.1];
        \fill (C) circle [radius=0.1];

        \fill (I) circle [radius=0.1] node [below] {$\ten_1 (-)$};
        \fill (II) circle [radius=0.1] node [below] {\hspace{15pt} $\ten_2 (+), \fives_1 (+), \eta (-)$};

        \draw ($(B)+(-.5,-1.5)$) .. controls ($(B)+(1,-1)$) and ($(B)+(0.5,-.2)$) .. ($(B)+(1.5,-.2)$) node (bulk) [right] {bulk};
        \path (bulk) node [below=0.1] {\hspace{30pt} $H (+), \fives_2 (-), \fives_3 (+)$}; 
    \end{tikzpicture}
    \caption{A picture of the $\Torb$ orbifold. The opposite sides of the parallelogram are identified. Three points I, I\!I, and I\!I\!I represent the fixed points. The localization of quarks $\ten_i$ and $\fives_i$, the Higgs $H$, and the scalars $\eta$ and $\delta$ are also shown with their $\Ztwo$ charges}
    \label{fig:T2/Z3}
\end{figure}

To obtain the diagonal mass matrix for the up-type quarks, $M_u$, we put $\ten$'s on different fixed points from each other as shown in Fig.\ref{fig:T2/Z3}.\footnote{We may argue that the presence of three fixed points is the primary origin of the three families of quarks and leptons. \cite{Watari:2002fd} places three $\fives$ at each fixed point. We discuss the details in Appendix~\ref{app:sixdimanomaly}. }
The Higgs field lives in the bulk, and we assume that its wave function varies at different fixed points, leading to the hierarchy of the Yukawa coupling in the mass matrix of the up-type quarks, $m_u : m_c : m_t \simeq \epsilon^2 : \epsilon : 1$ with $\epsilon \simeq 1/300$ \cite{Antusch:2013jca}. As to the off-diagonal terms, an exponential suppression, $e^{-M_* L}$, arises naturally between different fixed points if the length between two fixed points, $L$, is large enough. Such an exponential suppression is a crucial merit of using the extra dimensions. Here $M_* \simeq 10^{17}$ GeV is the six-dimensional Planck scale. 

We also need additional symmetry to engineer the down-type quark mass matrix. A minimal choice is $\Ztwo$, under which $\ten$'s can only have two possible charge assignments up to equivalence under permutation, $(+,+,+)$ or $(-,+,+)$, while only the latter could possibly lead to the desired texture in the down-type quark mass matrix.

We then want to engineer the three-zero texture of $M_d$ in Eq.\eqref{eq:Md1}. We start by putting all $\fives$'s in the bulk.
Since we fixed $\ten$'s to be $(-,+,+)$, we have four choices for $\fives$'s charges: $(+,+,+),\ (+,-,+), \ (+,-,-), \ (-,-,-) $, where permutation would lead to other textures in \cite{Tanimoto:2016rqy}.  Corresponding to the four charge choices, the possible textures are
\begin{equation}
 \left(\begin{array}{ccc}
0 & 0& 0 \\
\checkmark & \checkmark& \checkmark \\
\checkmark & \checkmark & \checkmark
\end{array}\right)\, , \ 
  \left(\begin{array}{ccc}
0 &  \checkmark &0  \\
\checkmark & 0&  \checkmark\\
\checkmark & 0&  \checkmark  
\end{array}\right)\, , \ \left(\begin{array}{ccc}
0 & \checkmark & \checkmark  \\
\checkmark & 0& 0\\
\checkmark & 0 & 0 
\end{array}\right)\, , \ \left(\begin{array}{ccc}
\checkmark & \checkmark& \checkmark \\
0 & 0& 0\\
0 & 0& 0 
\end{array}\right)\, ,
\end{equation}
where check marks denote the non-vanishing matrix elements under $\mathbb{Z}_2$, all of which are real due to CP symmetry. We further introduce the CP-violating singlet scalar field $\eta$, which carries the $\Ztwo$ odd charge. Notice that for the first and fourth charge choices, regardless of whether we place the $\eta $ filed in the bulk or at the fixed points, the matrix will not have the desired structure with a real determinant. For the second and third charge choices, placing the $\eta$ field at certain fixed points results in a matrix with a real determinant, although the matrix contains a complex phase. We will focus on the second charge choice for now.

To fully recover the mass matrix in Eq.\eqref{eq:Md1}, we place the CP-violating field $\eta$ on the second fixed point to introduce a complex phase in $(M_{d})_{22}$. To further break the $\mathbb{Z}_2$ symmetry, we introduce an additional real singlet scalar field $\delta$ with an odd $\mathbb{Z}_2$ charge at the third fixed point. Notice that now the mass matrix contains two zeros, which is a generic and unique prediction with one free parameter remaining to be tested in the future. However, since only the three-zero texture has been analyzed and compared with data in \cite{Tanimoto:2016rqy}, we require $\fives_1$ to be localized at the second fixed point to achieve the three-zero texture Eq.\eqref{eq:Md1}, as shown in Fig.\ref{fig:T2/Z3}. We summarize the $\Ztwo$-charge assignment in Tab.\ref{tab:Ztwo_charge}.

\vspace{\baselineskip}
\begin{table}[h]
    \centering
    \begin{tabular}{|wc{20pt}||wc{20pt}|wc{20pt}|wc{20pt}|wc{20pt}|wc{20pt}|wc{20pt}|wc{20pt}|wc{20pt}|wc{20pt}|}
        \hline
        & $\ten_1$ & $\ten_2$ & $\ten_3$ & $\fives_1$ & $\fives_2$ & $\fives_3$ & $H$ & $\eta$ & $\delta$  \\ \hline
        $\Ztwo$ & $-$ & $+$ & $+$ & $+$ & $-$ & $+$ & $+$ & $-$ & $-$ \\ \hline
    \end{tabular}    
    \caption{$\Ztwo$ charge for each particle. }
    \label{tab:Ztwo_charge}
\end{table}

We now discuss the spontaneous symmetry breaking of the $\Ztwo$ symmetry and the CP symmetry by a complex singlet scalar $\eta$ and a real singlet scalar $\delta$ in detail. We assume their potentials at the fixed point I\!I and I\!I\!I as
\begin{equation}
    V_\mathrm{I\!I}(\eta) = -\mu_1 \eta \eta^\dagger - \mu_2 (\eta^2 + \eta^{\dagger 2}) + \lambda_1 (\eta \eta^\dagger)^2 + \lambda_2 (\eta^4 + \eta^{\dagger 4})\ ,
    \label{eq:VII}
\end{equation}
and
\begin{equation}
    V_\mathrm{I\!I\!I}(\delta) = \lambda_3 (\delta^2 - v_\delta^2)^2\ ,
\end{equation}
where all parameters $\mu_{1,2}$, $\lambda_{1,2,3}$, and $v^2_\delta$ are real due to the CP symmetry. We assume that $\eta$ is heavy enough to ignore the Higgs coupling. 
With the given potentials, $\eta$ acquires a complex VEV, which accounts for the complex phase in $(M_{d})_{22}$. This, along with $\braket{\delta} = v_\delta$, breaks the $\Ztwo$ symmetry.
 
Saving the details of a possible UV completion model in Appendix \ref{sec:UV_completion}, the effective operator for $(M_{d})_{22}$ takes the form 
\begin{align}
    \left(\frac{c_1}{\Lambda} \vev{\eta} + \frac{c_2}{\Lambda} \vev{\eta^\dagger}\right) \ten_2 H \fives_2\ ,
    \label{eq:Md22}
\end{align}
where $\Lambda \simeq \braket{\eta}$ is the cutoff scale of the CP-symmetry breaking, and the number in the parenthesis is the coefficient $b e^{-i\phi}$ in Eq.\eqref{eq:Md1} at tree level. These coefficients $c_{1,2}$ are proportional to the Higgs wave function at the fixed point I\!I, which is suppressed by $\epsilon$. 

The CP-violating field could also introduce a complex phase in $(M_{d})_{21}$ through a coupling $\eta^2\ten_2 H \fives_1 $ for example, which further introduces a complex phase in the determinant.\footnote{Similar argument holds as well for $(M_{d})_{23}$. } However, this term is forbidden in the UV-completion model discussed in Appendix \ref{sec:UV_completion}.
Therefore, the coefficients should be suppressed by the six-dimensional Planck scale $M_*$. To have physical angle $|\bar\theta| \lesssim 10^{-10}$, $\Lambda/M_* \lesssim 10^{-5}$ has to be satisfied, and we predict the CP-violation scale as $\Lambda\lesssim 10^{12}$ GeV. 
For simplicity, we take the cutoff scale $\Lambda\simeq \vev{\eta} \simeq \vev{\delta}$. An illustrative model of the corresponding high-energy theory is given in Appendix \ref{sec:UV_completion}.

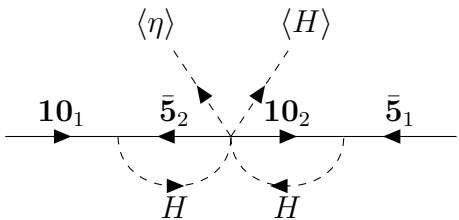
\begin{figure}
    \centering
    \begin{tikzpicture}[>={Latex[]}]
    \begin{feynman}
       \vertex (a) at (-3,0);
    \vertex (b) at (-1.5,0);
    \vertex (c) at (0,0);
    \vertex (d) at (1.5,0);
    \vertex (e) at (3,0);
    \vertex (o1) at (-1,1.5) {\(\braket{\eta} \)};
    \vertex (o2) at (1,1.5)  {\(\braket{H} \)};
        \diagram*{
        (a) -- [fermion, edge label=\(\ten_1 \)] (b),
        (b) -- [anti fermion, edge label=\(\fives_2 \)] (c),
        (c) -- [fermion, edge label=\(\ten_2 \)] (d),
        (d) -- [anti fermion, edge label=\(\fives_1 \)] (e),
        (b) -- [charged scalar, half right, edge label'=\(H \)] (c),
        (c) -- [anti charged scalar, half right, edge label'=\(H \)] (d),
        (c) -- [charged scalar] (o1),
        (c) -- [charged scalar] (o2),
    };
 \end{feynman}        
    \end{tikzpicture}
    \caption{A two-loop diagram that can contribute to the complex phase in $(M_d)_{11}$.}
    \label{fig:2loop}
\end{figure}

The effective theory below the cutoff scale can be obtained by the integration of heavy particles such as the heavy $\eta$ and $\delta$ (and the heavy Higgs particles present in the UV model given in Appendix \ref{sec:UV_completion}). It is the SM Lagrangian with higher-order interaction terms such as 
\begin{equation}
\label{eq,higgsinteraction}
   g \frac{\braket{\eta}}{\Lambda} \ten_2 H \fives_2 \frac{H^\dagger H}{\Lambda^2} + \mathrm{h.c.}\ ,
\end{equation}
where $g$ is a coefficient whose explicit form is given in Eq.\eqref{eq,effectivecoeff}.
As a result, the physical vacuum angle ${\bar \theta}$ vanishes at tree level. This means that the loop correction from the SM sector to the vacuum angle shift only occurs at four-loop level, where it is strongly suppressed to $|{\bar \theta}| < 10^{-16}$ as demonstrated in \cite{Ellis:1978hq}. However, effective operators like Eq.\eqref{eq,higgsinteraction} introduce additional complex phases at two-loop level, as shown in Fig.\ref{fig:2loop} for example. Without delving into the details of the loop computation, we estimate the value of this diagram through tree-level values and find that the correction to $(M_d)_{11}$ is proportional to
\begin{equation}
    \Delta_{11} \sim \frac{\vev{\eta}}{\Lambda} \frac{g}{\Lambda^2 }\left(\frac{1}{16 \pi^2}\right)^2 \Lambda^2 \frac{a}{v} \frac{a^{\prime}}{v} v \sim \frac{\vev{\eta}}{\Lambda} g v \times 10^{-13}  \ .
\end{equation}
where $(1/16 \pi^2)^2\Lambda^2$ is the loop factor, the coefficients $a,a'$ are taken from Tab.\ref{tab:params}, and $v = 246$ GeV is the SM-Higgs VEV. Hence, the contribution of $ \Delta_{11}$ to $\bar{\theta}$ is roughly estimated as
\begin{align}
    \bar{\theta} = \mathrm{Arg}\det M_d 
    \sim \frac{\mathrm{Im}[\Delta_{11} (b e^{-i\phi} d-cc')] }{aa'd}
    \sim g  \times 10^{-9} \ .
\end{align}
As discussed in Appendix \ref{sec:UV_completion}, the coupling constants in 4D are described by the overlap with the Higgs wave function at each fixed point. Thus, we expect that $g\sim\mathcal{O}(\epsilon^3)$, and the above two-loop diagram is sufficiently suppressed, $|\bar\theta| \lesssim 10^{-16}$. We therefore achieve a consistent model solving the strong CP problem through the three-zero texture of the down-type quark mass matrix.

\section{Conclusions and discussion}
\label{sec:conclusion}

In this paper, we propose a solution to the strong CP problem through the three-zero texture of the down-type quark mass matrix at the effective field theory level. Utilizing the extra dimensions, and their compactification to the orbifold $\Torb$, we construct a model that naturally leads to this three-zero texture.

In this framework, we place $\ten$'s on three fixed points to obtain a diagonal up-type quark mass matrix. The mass hierarchy of the up-type quark is related to the Higgs wave function at each fixed point. By utilizing the additional $\mathbb{Z}_2$ symmetry, as summarized in Tab.\ref{tab:Ztwo_charge}, we place the down-type quarks in the torus bulk and introduce a CP-violating scalar $\eta$ and a real scalar $\delta$, both of which are $\mathbb{Z}_2$ odd and break the additional symmetries by acquiring VEVs. This results in a two-zero texture for the down-type quark mass matrix.  Remarkably, the determinant of this two-zero texture is real, and the theory does not suffer from the strong CP problem. To explain the three-zero texture that fits the data \cite{Tanimoto:2016rqy}, we localize $\fives_1$ at the second fixed point. 

The three-zero texture of the down-type quark mass matrix results in a vanishing physical vacuum angle, $\bar\theta = 0$, at tree level, and hence, the loop corrections from the SM interactions are suppressed at four-loop level as shown in \cite{Ellis:1978hq}. Thus, we only need to consider corrections from non-SM interactions arising from integrating out heavy particles. Due to the hierarchy of the Higgs wave function at different fixed points, these corrections to the physical vacuum angle are suppressed to $|\bar\theta| \lesssim 10^{-16}$ at two-loop level. Our model solves the strong CP problem without radiative corrections, clearly distinguishing it from the Nelson-Barr model, which experiences large loop corrections to ${\bar \theta}$ at one-loop level \cite{Dine:2015jga}.

We would like to emphasize that the localization of the CP-violating field $\eta$ at the fixed point I\!I is crucial. Otherwise, $(M_d)_{11}$ can also contain $\braket{\eta}$, which would generate the complex phase of the determinant of $M_d$, and ${\bar \theta}$ would no longer be vanishing. Such a localization of the CP-violating field can be realized only in the higher-dimensional spacetime.

In this paper, we focused on Eq.\eqref{eq:Md1} as one example of a three-zero texture quark mass matrix that can fit the data while having a real determinant. Extending this to other mass textures, as explored in \cite{Tanimoto:2016rqy}, is straightforward. A common feature among these models is the requirement of a higher-dimensional theory to achieve real determinants for both the up-type and down-type quark matrices. Additionally, the distance $L$ between the fixed points must be sufficiently large, with $M_* L \gtrsim 30$, to satisfy the experimental constraint $|\bar \theta| < 10^{-10}$. We do not address the high-energy fundamental theory underlying this phenomenological model, nor do we calculate the Higgs wave function in the bulk. Instead, we assume the hierarchy of quark masses as a given fact, which is essential for achieving suppressed loop corrections. It is worth noting that our model naturally predicts a two-zero texture in the down-type quark mass matrix, which should be tested and constrained by experimental data in future work.

It is straightforward to extend our mechanism to the lepton sector with three right-handed neutrinos (RHNs). The large Majorana masses of the RHNs are crucial for explaining the tiny Majorana masses of the left-handed neutrinos via the seesaw mechanism \cite{Yanagida:1979as, Gell-Mann:1979vob, Glashow:1979nm, Minkowski:1977sc}. Decays of the RHNs in the early universe produce a lepton asymmetry, which is then converted into a baryon number asymmetry through the leptogenesis mechanism \cite{Fukugita:1986hr}. Since the only source of CP violation in our model is the complex phase of the VEV of the $\eta$ field, the sign of the baryon asymmetry \cite{Frampton:2002qc} may be related to CP-violating phases at low energies, such as those in the CKM quark mixing matrices and the Pontecorvo-Maki-Nakagawa-Sakata neutrino mixing matrices. The details of these processes will be discussed in future publications.

\section*{Ackowledgments}
We would like to thank Morimitsu Tanimoto for valuable information on the quark mass matrices and encouragement. We would acknowledge Kazuya Yonekura, Taizan Watari and Yoshiharu Kawamura for discussion on the $\Torb$ orbifold compactification.
This work is supported by 
JSPS Grant-in-Aid for Scientific Research
Grants No.\,24H02244, the National Natural Science
Foundation of China (12175134)
and World Premier International Research Center
Initiative (WPI Initiative), MEXT, Japan. 
RO was supported by Forefront Physics and Mathematics Program to Drive Transformation (FoPM), a World-leading Innovative Graduate Study (WINGS) Program, the University of Tokyo,
JSPS KAKENHI Grant Number JP24KJ0838, 
and JSR Fellowship, the University of Tokyo.

\appendix
\section{Origin of the $\fives$'s in the bulk }
\label{app:sixdimanomaly}
In this section, we discuss a possible origin of the bulk $\fives$'s in the six-dimensional spacetime following notation in \cite{Goto:2017zsx}. We introduce an anomaly-free Dirac fermion consisting of the positive chirality $\Psi_+$ and the negative chirality $\Psi_-$ in six dimensions that are obtained from the projection,
\begin{equation}
    \Psi_+ = \left(\begin{array}{cc}
\frac{1-\gamma_5}{2} & 0 \\
0 & \frac{1+\gamma_5}{2}
\end{array}\right) \Psi=\binom{\psi_{+L}}{\psi_{+R}} \ ,
\end{equation} 
and \begin{equation}
    \Psi_- = \left(\begin{array}{cc}
\frac{1+\gamma_5}{2} & 0 \\
0 & \frac{1-\gamma_5}{2}
\end{array}\right) \Psi=\binom{\psi_{-R}}{\psi_{-L}} \ . 
\end{equation} 

We denote the six-dimensional coordinates with the orbifold $\Torb$ as $x= (x_\mu,z)$, where $\mu = 0,1,2,3$ are the four-dimensional spacetime coordinates, and $z$ is the complex two-dimensional coordinate. There exists a $\mathbb{Z}_3$ symmetry under which the coordinates rotate as
\begin{equation}
    \sigma : (x_\mu,z)\mapsto (x_\mu,\sigma z) \hspace{10pt}
    \text{where} \hspace{10pt} \sigma z \equiv \omega z = \omega^{-2} z \, , \ \omega \equiv e^{\frac{2\pi i }{3}}\ .
\end{equation}
Under this symmetry, the fermions transform as
\begin{equation}
    \sigma: \psi_{+L(R)} (x_\mu,z) \mapsto \eta_{+L(R)} \psi_{+L(R)} (x_\mu,\sigma z) \ ,
\end{equation}
and 
\begin{equation}
    \sigma: \psi_{-R(L)} (x_\mu,z) \mapsto \eta_{-R(L)} \psi_{-R(L)} (x_\mu,\sigma z) \ ,
\end{equation}
where $\eta_{+R} = \omega \eta_{+L}$ is the phase picked up by the right-handed chiral fermion $\psi_{+R}$. Similarly, $\eta_{-L} = \omega \eta_{-R}$. For a mass term $\bar\Psi \Psi$ to be allowed in principle, the phases are related through $\eta_{+L} =\eta_{-R} $. We are left with one degree of freedom to choose  $\eta_{+L} =\omega^{-1}$,\footnote{We would like to thank Yoshiharu Kawamura for an intensive discussion and valuable comments on the transformation properties of the six-dimensional fermions under the orbifold $\mathbb{Z}_3$ symmetry. }  
and $\psi_{+R}$ and $\psi_{-L}$ are invariant under this $\mathbb Z_3$ symmetry and remain massless after the orbifold compactification.

Now, we assign the $\SU(5)$ representation of the six-dimensional Dirac fermion as $\psi(\five)$, which contains a pair of the massless four-dimensional fermions, $\psi_{+R}(\five)$ and $\psi_{-L}(\five)$. Now, we introduce three sets of the Dirac fermions in the bulk\footnote{One might start with an anomaly-free pair of $\fives'$ and $\ten$ at each fixed point, which appears to be a more elegant model. However, for large distances between the fixed points, there is no flavor mixing. }
to make the model realistic, and hence we have three sets of massless four-dimensional fermions
after the orbifold compactification. We will drop the $\pm$ sign of the four-dimensional fermions for the rest of this appendix.

In addition to the above fermions, we introduce a pair of anomaly-free four-dimensional fermions at each fixed point, namely $\psi_{R}'(\five)$ and $\psi_{L}(\ten)$. The left-handed fermions, $\psi_{L}(\five)$, gain mass at the fixed point through the interaction $M{\overline {\psi_{R}'}(\five)} \psi_{L}(\five)$, resulting in three massless right-handed fermions, $\psi_{R}(\five)$, in the bulk, which correspond to the $\fives$'s in the main context. We later restrict $\fives_1$ at the second fixed point to get the three-zero texture.\footnote{We can also start from the localized $\fives_1$ on the second fixed point, and introduce two pairs of fermions in the bulk to cancel anomaly. } 
The three massless left-handed fermions, $\psi_{L}(\ten)$, are the $\ten$'s in the main context. They will gain mass through the Higgs mechanism. Therefore, we obtain anomaly-free chiral fermions in 6D.

\section{A UV complete model}
\label{sec:UV_completion}

In this section, we discuss a UV completion for the effective operators $\frac{\vev{\eta}}{\Lambda} \ten_2 H \fives_2$, $\frac{\vev{\eta^\dagger}}{\Lambda} \ten_2 H \fives_2$, and $\frac{\vev{\delta}}{\Lambda} \ten_3 H \fives_2$ discussed in Eq.\eqref{eq:Md22}. 

First, we start with the UV completion for $\frac{\vev{\eta}}{\Lambda} \ten_2 H \fives_2$ on the fixed point I\!I.
We introduce a new heavy Higgs boson $\Htwo$ with an odd $\Ztwo$ charge on the fixed point I\!I. This heavy Higgs boson introduces the following renormalizable interactions 
\begin{align}
    &M_{\mathrm{I\!I}}^2 \Htwo^\dagger \Htwo + Y (\ten_2 \Htwo \fives_2 + \mathrm{h.c.})
     \notag \\
    & + \alpha_1 (\Htwo^\dagger H \eta + H^\dagger \Htwo \eta^\dagger) + \alpha_2 (H^\dagger \Htwo \eta + \Htwo^\dagger H \eta^\dagger) \notag \\
    &+ \beta_1 (\Htwo^\dagger \Htwo)^2 + \beta_2 H^\dagger H \Htwo^\dagger \Htwo + \beta_3 [ (H^\dagger \Htwo)^2 + (\Htwo^\dagger H)^2 ]\ ,
    \label{eq:H_II}
\end{align}
where $M_\mathrm{I\!I} \simeq \Lambda$ and $\alpha_{1,2}$ are dimensionful parameters, and $Y$ and $\beta_{1,2,3}$ are dimensionless parameters.\footnote{There are other interaction terms such as $H^\dagger H \eta^\dagger \eta$ and $\Htwo^\dagger \Htwo \eta^\dagger \eta$, but they only contribute to the mass renormalization of $H$ and $\Htwo$.}
All of these parameters are real due to the CP symmetry.

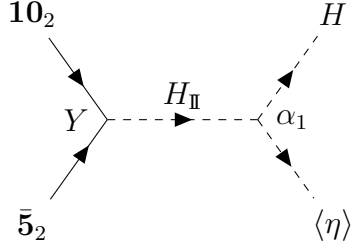
\begin{figure}
    \centering
    \begin{tikzpicture}[>={Latex[]}]
        \begin{feynman}
            \vertex (A) at (-2,1.4) {$\ten_2$};
            \vertex (B) at (-2,-1.4) {$\fives_2$};
            \vertex (O1) at (-1,0);
            \vertex (O2) at (1,0);
            \vertex (C) at (2,1.4) {$H$};
            \vertex (D) at (2,-1.4) {$\vev{\eta}$};

            \node[left=0.1cm] at (O1) {$Y$};
            \node[right=0.1cm] at (O2) {$\alpha_1$};
            
            \diagram*{
                (A) -- [fermion] (O1) -- [anti fermion] (B);
                (O1) -- [charged scalar, edge label={$\Htwo$}] (O2) -- [charged scalar] (C);
                (O2) -- [charged scalar] (D);
            };
        \end{feynman}
    \end{tikzpicture}
    \caption{A Feynman diagram for the $\Htwo$ exchange.}
    \label{fig:Hpex}
\end{figure}

Below the energy scale $\Lambda$, the $\Htwo$ exchange diagram Fig.\ref{fig:Hpex} induces the effective operator 
\begin{align}
    \frac{Y \alpha_1}{M_{\mathrm{I\!I}}^2} \vev{\eta} \ten_2 H \fives_2
    \simeq Y\alpha_1\frac{\vev{\eta}}{\Lambda^2} \ten_2 H \fives_2\ .
\end{align}
This gives a complex phase to the down-type quark mass matrix at tree level: specifically in terms of Eq.\eqref{eq:Md22}, $c_1\sim \frac{Y\alpha_1}{\Lambda}$.
Another effective operator $\frac{\vev{\eta^\dagger}}{\Lambda} \ten_2 H \fives_2$ can be obtained in the same way with replacing $\alpha_1$ with $\alpha_2$.

\begin{figure}
    \centering
    \begin{tikzpicture}[>={Latex[]}]
    \begin{feynman}
        \vertex (O1) at (-5,0);
        \vertex[left=1.7cm of O1] (a1) {$\ten_2$};
        \vertex[right=1.7cm of O1] (b1) {$\fives_2$};
        \vertex[above=1.5cm of O1] (G1);
        \vertex[left=0.9cm of G1] (A1);
        \vertex[above=1.5cm of G1] (A2);
        \vertex[right=0.9cm of G1] (A3);
        \vertex (c1) at ($(a1)+(0,1.5)$) {$H$};
        \vertex[above=2.3cm of a1] (d1) {$\vev{\eta}$};
        \vertex(h1) at ($(b1)+(0,1.5)$) {$H$};
        \vertex[above=2.3cm of b1] (g1) {$\vev{\eta}$};
        \vertex (e1) at ($(A2)+(-1.5,1)$) {$H^\dagger$};
        \vertex (f1) at ($(A2)+(1.5,1)$) {$\vev{\eta}$};

        \node[below] at (O1) {$Y$};
        \node[above right] at (G1) {$\beta_1$};
        \node[below] at (A1) {$\alpha_1$};
        \node[above=0.1cm] at (A2) {$\alpha_2$};
        \node[below] at (A3) {$\alpha_1$};
        
        \diagram{
            (a1) -- [fermion] (O1) -- [anti fermion] (b1);
            (O1) -- [charged scalar] (G1) -- [charged scalar] (A1) -- [charged scalar] (c1);
            (A1) -- [charged scalar] (d1);
            (e1) -- [charged scalar] (A2) -- [charged scalar] (G1) -- [charged scalar] (A3) -- [charged scalar] (g1);
            (A2) -- [charged scalar] (f1);
            (A3) -- [charged scalar] (h1);
        };
        
        \vertex (O2) at (0,0);
        \vertex[left=1.7cm of O2] (a2) {$\ten_2$};
        \vertex[right=1.7cm of O2] (b2) {$\fives_2$};
        \vertex[above=1.5cm of O2] (G2);
        \vertex[above=1.5cm of G2] (B1);
        \vertex (c2) at ($(a2)+(0,1.5)$) {$H^\dagger$};
        \vertex (d2) at ($(b2)+(0,1.5)$) {$H$};
        \vertex (e2) at ($(B1)+(-1.5,1)$) {$H$};
        \vertex (f2) at ($(B1)+(1.5,1)$) {$\vev{\eta}$};

        \node[below] at (O2) {$Y$};
        \node[above right] at (G2) {$\beta_2$};
        \node[above=0.1cm] at (B1) {$\alpha_1$};
        
        \diagram{
            (a2) -- [fermion] (O2) -- [anti fermion] (b2);
            (O2) -- [charged scalar] (G2) -- [charged scalar] (B1) -- [charged scalar] (e2);
            (c2) -- [charged scalar] (G2) -- [charged scalar] (d2);
            (B1) -- [charged scalar] (f2);
        };

        \vertex (O3) at (5,0);
        \vertex[left=1.7cm of O3] (a3) {$\ten_2$};
        \vertex[right=1.7cm of O3] (b3) {$\fives_2$};
        \vertex[above=1.5cm of O3] (G3);
        \vertex[above=1.5cm of G3] (C1);
        \vertex (c3) at ($(a3)+(0,1.5)$) {$H$};
        \vertex (d3) at ($(b3)+(0,1.5)$) {$H$};
        \vertex (e3) at ($(C1)+(-1.5,1)$) {$H^\dagger$};
        \vertex (f3) at ($(C1)+(1.5,1)$) {$\vev{\eta}$};

        \node[below] at (O3) {$Y$};
        \node[above right] at (G3) {$\beta_3$};
        \node[above=0.1cm] at (C1) {$\alpha_2$};
        
        \diagram{
            (a3) -- [fermion] (O3) -- [anti fermion] (b3);
            (O3) -- [charged scalar] (G3) -- [anti charged scalar] (C1) -- [anti charged scalar] (e3);
            (c3) -- [anti charged scalar] (G3) -- [charged scalar] (d3);
            (C1) -- [charged scalar] (f3);
        };
    \end{feynman}
    \end{tikzpicture}
    \caption{Feynman diagrams that lead to the effective operator Eq.\eqref{eq:higher_II} with $n=1$ and $l=0$. The internal propagators indicate the $\Htwo$ propagators, which give $1/M_\mathrm{I\!I}^2 \sim 1/\Lambda^2$ after integrating out $\Htwo$.}
    \label{fig:HdH10H5}
\end{figure}
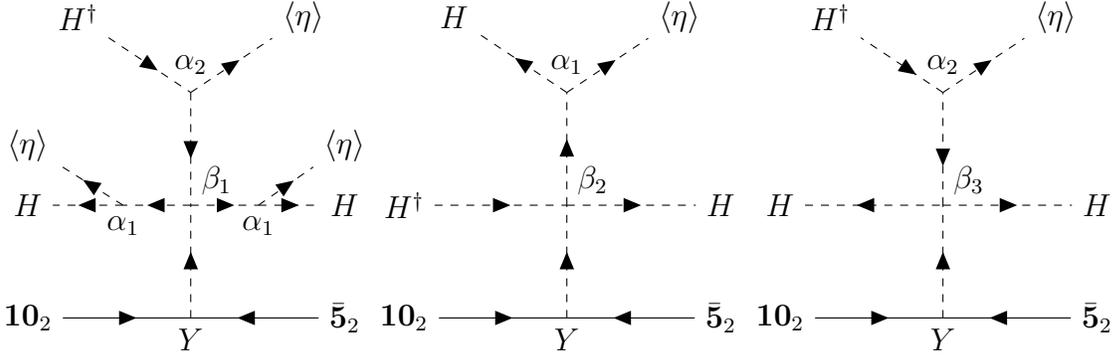

Similarly, using Higgs four-point interactions in Eq.\eqref{eq:H_II}, we can obtain the following higher-dimensional operators
\begin{align}
   \frac{\vev{\eta}^m \vev{\eta^\dagger}^{l}}{\Lambda^{2n+m+l}} (H^\dagger H)^n \ten_2 H \fives_2\ ,
    \label{eq:higher_II}
\end{align}
where the coefficient is omitted.
For example, in the case of $n=1$ and $l=0$, we show three diagrams that introduce such operators in Fig.\ref{fig:HdH10H5}.
After integrating out $\Htwo$, these diagrams give a dimension-six effective operator
\begin{align}
\label{eq,effectivecoeff}
   Y 
   \left[
    \frac{\alpha_1^2 \alpha_2 \beta_1}{\Lambda^3} \frac{\vev{\eta}^3}{\Lambda^3} 
    + \frac{\alpha_1 \beta_2 + \alpha_2 \beta_3}{\Lambda} \frac{\vev{\eta}}{\Lambda} 
   \right] 
   \frac{H^\dagger H}{\Lambda^{2}} \ten_2 H \fives_2\ .
\end{align}
Since we assume that $\vev{\eta}$ and $\vev{\eta^\dagger}$ are complex numbers of $\mathcal{O}(\Lambda)$, Eq.\eqref{eq:higher_II} gives effective operators 
\begin{align}
    \mathcal{O}^{(n)}_\mathrm{I\!I} \equiv \frac{c^{(n)}_\mathrm{I\!I}}{\Lambda^{2n}} (H^\dagger H)^n \ten_2 H \fives_2\ ,
\end{align}
where $c^{(n)}_\mathrm{I\!I}$ is a complex coefficient.

The higher-order operator with $n=1$ is the most dangerous since it introduces the largest radiative corrections on the ${\bar \theta}$ at two-loop level as shown in the text. However, the coupling constants $\alpha_{1,2}$ and $\beta_{2,3}$ depend on the model, and there is, in fact, an interesting phenomenological model where those coupling constants are all suppressed. Notice that the $\alpha_1/\Lambda$ must be already as small as $10^{-3}$ to reproduce the CKM matrix, provided $Y\simeq 1$; see Tab.\ref{tab:params}. Thus, all dangerous higher-order operators are strongly suppressed in the model. We take such a model in this paper and the effective theory below the scale $\Lambda$ is well described by the SM with ${\bar \theta} =0$ at tree level. This is the situation that Ellis and Gaillard considered in \cite{Ellis:1978hq} and all radiative corrections on ${\bar \theta}$ are sufficiently suppressed. We will explain this model at the end of this section.

On the fixed point I\!I\!I, we introduce another new Higgs boson $\Hthr$ which has the following interaction terms
\begin{align}
\label{eq,HH3interaction}
    &M_{\mathrm{I\!I\!I}}^2 \Hthr^\dagger \Hthr + Y' (\ten_3 \Hthr \fives_2 + \mathrm{h.c.})
    + \alpha' (\Hthr^\dagger H + H^\dagger \Hthr) \delta \notag\\
    &+ \beta'_1 (\Hthr^\dagger \Hthr)^2 + \beta'_2 H^\dagger H \Hthr^\dagger \Hthr + \beta'_3 [ (H^\dagger \Hthr)^2 + (\Hthr^\dagger H)^2 ]\ ,
\end{align}
Here again, all parameters, $M_\mathrm{I\!I\!I}$, $Y'$, $\alpha'$, and $\beta'_{1,2,3}$, are real.
As in the case of the fixed point I\!I, these interactions give an effective operator
\begin{align}
    \frac{Y'\alpha'}{M_{\mathrm{I\!I\!I}}^2} \vev{\delta} \ten_3 H \fives_2
    \simeq Y'\alpha' \frac{\vev{\delta}}{\Lambda^2} \ten_3 H \fives_2\ .
\end{align}
The following higher-dimensional operators can be obtained in a similar way,
\begin{align}
\label{eq,B8}
    \frac{\vev{\delta}^{m}}{\Lambda^{2n+m}} (H^\dagger H)^n \ten_3 H \fives_2\ ,
\end{align}
which leads to
\begin{align}
    \mathcal{O}^{(n)}_\mathrm{I\!I\!I} \equiv \frac{c^{(n)}_\mathrm{I\!I\!I}}{\Lambda^{2n}} (H^\dagger H)^n \ten_3 H \fives_2\ ,
\end{align}
where $c^{(n)}_\mathrm{I\!I\!I}$ is a real coefficient since $\vev{\delta}$ is real.

Our phenomenological model is motivated by the observed large mass hierarchy for the up-type quarks, that is, $m_u : m_c : m_t \simeq \epsilon^2 : \epsilon : 1$ with the $\epsilon \simeq 1/300$ \cite{Antusch:2013jca}. Such a large hierarchy seems unnatural, since the three fixed pints, I, I\!I, and I\!I\!I, are equivalent in the present $\Torb$ orbifold. Recall that the quark mass is given by $m_i = y_i\times v$, where the $y_i$ and $v$ are a Yukawa coupling constant of the Higgs $H$ to the quark $\ten_i$ and the VEV of the $H$, respectively. Now, the Yukawa coupling constant $y_i$ depends on the overlapping of the wave functions of the Higgs, $\Psi(H)$, and the $\ten_i$ at the fixed point $i$ in the two-dimensional bulk as discussed in \cite{Arkani-Hamed:1999ylh}. Thus it is quite natural to consider the ratio of the wave functions at the fixed points to satisfy $|\Psi(H)|_{\rm I} : |\Psi(H)|_{\rm I\!I}: |\Psi(H)|_{\rm I\!I\!I} \simeq \epsilon^2 : \epsilon : 1$. Here, $|\Psi(H)|_i$ is the size of the wave function at the fixed point $i=$ I, I\!I, I\!I\!I.\footnote{The wave function of the Higgs in the two-dimensional bulk, $\Psi(H)$, may be determined by interactions of unknown bulk fields and boundary conditions. If the Higgs has a non-flat wave function, it has a large four-dimensional effective mass of $\mathcal{O}(\Lambda^2)$ from curvature derivative. However, after $\delta$ obtains a VEV, there is mixing between $H$ and $H_{\rm I\!I\!I}$ in Eq.\eqref{eq,HH3interaction}, which can introduce a mass eigenstate with a negative mass square at the same magnitude of order. The detailed cancellation may give the light SM-Higgs mass that we observe.} 

If this is the case, all coupling constants of the dangerous operators discussed above are strongly suppressed by a factor $\epsilon^2 \sim 10^{-5}$. For example, the coefficient of the operator, $\frac{\vev{\eta}}{\Lambda} \frac{H^\dagger H}{\Lambda^2} \ten_2 H \fives_2$, is suppressed by $\mathcal{O}(\epsilon^3)$. It is easy to see that the radiative corrections from this operator are not dangerous at all, taking into account the two-loop effect and the multi-products of other Yukawa coupling. This is the reason why we neglect the higher-order terms in this paper. The effect of the higher-order terms with $\delta$ in Eq.\eqref{eq,B8} is not problematic since $\delta$ is real, as long as $(M_d)_{11}$ and $(M_d)_{31}$ vanish. We can easily confirm this at two-loop level since $\ten_3$ is decoupled from $\fives_1$. However, it should be noted here that the small mass, $\mathcal{O}(1)$ MeV, of the up quark might be required by the anthropic principle \cite{Weinberg:1987dv,Damour:2007uv} and hence the ratio of $|\Psi(H)|_{\rm I} /|\Psi(H)|_{\rm I\!I\!I} $ can be larger than $\epsilon^2$. This point might be important when we discuss the mass matrix of the down-type quarks.

As for the mass matrix of the down-type quarks, $M_d$, it is not easy to predict the mass hierarchy in our phenomenological model, since the origins of each $\fives_i$ are not equivalent. In particular, $\fives_2$ and $\fives_3$ live in the two-dimensional bulk, and their wave function profiles remain unknown. However, an interesting observation can be made: if the two wave function profiles in the bulk are the same, we can predict that the absolute value in the $(2,2)$ element is almost equal to that in the $(2,3)$ element, as well as the absolute value of the $(3,2)$ and $(3,3)$ elements. Remarkably, this prediction is consistent with the result in \cite{Tanimoto:2016rqy}. We leave the detailed discussion of the wave functions of $\fives_i$ for future work, as it is beyond the scope of the present paper.

\bibliographystyle{utphys}
\bibliography{ref}

\end{document}